# Title

Thermal Reconversion of Oxidised Lead White in Mural Paintings *via* a Massicot Intermediate

# Authors

Théa de Seauve[a]*, Sophie Bosonnet[b], Olivier Grauby[c], Alexandre Semerok[d], Vincent Detalle[e], Jean-Marc Vallet[a]

[a] Centre Interdisciplinaire de Conservation et Restauration du Patrimoine (CICRP), 21 rue Guibal, 13003 Marseille, France

[b] Université Paris-Saclay, CEA, Service de la Corrosion et du Comportement des Matériaux dans leur Environnement, 91191, Gif-sur-Yvette, France

[c] Aix-Marseille Université, CNRS – UMR 7325 CINaM, campus de Luminy, Case 913, 13288 Marseille Cedex 9, France

[d] Université Paris-Saclay, CEA, Service d'Études Analytiques et de Réactivité des Surfaces, 91191, Gif-sur-Yvette, France

[e] C2RMF, Palais du Louvre – Porte des Lions, 14, quai François Mitterrand, 75001 Paris, France

* Corresponding author: thea.de-seauve@cicrp.fr

# Abstract

Lead white is the most ancient and common white pigment used in mural paintings. However, it tends to blacken with time due to its oxidation to plattnerite ($\beta$-$PbO_2$). Chemical treatments were used but they can put the pictorial layers supports at risks. Hereby we address the possibility of thermally reconverting black plattnerite to white lead carbonates *via* a massicot ($\beta$-PbO) intermediate, with a view to developing a restoration procedure using continuous wave laser heating. We first investigated the conditions (temperature, time, and environment) in which pure powders react, before studying mural painting samples. Experiments were made in ovens and TGA and XRD and SEM characterization were achieved. Litharge (α-PbO) and massicot were obtained from plattnerite respectively between 564 and 567 °C and at 650 °C. Lead carbonates (cerussite, hydrocerussite and plumbonacrite) formed from massicot in wet $CO_2$ below 100 °C in a few hours. Lastly, when heating plattnerite based mural painting samples, lead species reacted with binders and mortar, yielding massicot, plumbonacrite but also lead silicate and calcium lead oxides. This demonstrates the viability of thermal reconversion of darkened lead in mural, while raising concerns about the formation of several lead species by reaction with mural painting constituents.

## Highlights

- Litharge (α-PbO) can be obtained by heating plattnerite (β-$PbO_2$) for 7 hours between 564 and 567 °C
- Massicot (β-PbO) can be obtained by heating plattnerite (β-$PbO_2$) for 7 hours at 650 °C.
- No thermal transformations of massicot (α-PbO) or litharge (β-PbO) were observed between 250 °C and 280 °C up to *ca.* 1 day in $CO_2$ and $CO_2/H_2O$
- Thermal transformation of massicot (α-PbO) in cerussite ($PbCO_3$), hydrocerussite ($2PbCO_3 \cdot Pb(OH)_2$) and plumbonacrite ($3PbCO_3 \cdot Pb(OH)_2 \cdot PbO$) was observed between 20 °C and 90 °C in a few hours in $CO_2$ in presence of water.
- Plumbonacrite, lead silicate $Pb_3SiO_5$, and calcium lead oxide $Ca_2PbO_4$ form when heating a plattnerite-based lime mural painting sample at 650 °C for one hour.

## Intro

Lead white is the most ancient and common white pigment used in mural paintings [1], [2]. However, this mixture of cerussite ($PbCO_3$) and hydrocerussite ($2PbCO_3 \cdot Pb(OH)_2$) tends to blacken with time, which has been known for a long time [2]–[4], due to its oxidation to plattnerite (β-$PbO_2$) [5]–[28].

Chemical treatment were used in order to achieve the reconversion[1] to lead white [29], [30], but hey can put the pictorial layers supports at risks [14]–[17]. The possibility to induce the reconversion reaction by continuous wave (CW) laser heating, with the aim of developing an intuitive, cheap and ready-to-use restoration device was previously addressed. We demonstrated the possibility of reconverting oxidised red lead by CW laser heating, but suggested that lead white thermal reconversion was possible only *via* an intermediate compound [31]–[33]. Hereby we address the possibility of thermally reconverting black plattnerite to white lead carbonates *via* a massicot (β-PbO) intermediate.

Plattnerite is thermally decomposed into PbO via several intermediate compounds [47]–[54]. Litharge (α-PbO) appears at lower temperatures than massicot, which is the final stage of $PbO_2$ decomposition [49]–[54]. Risold *et al.*'s review indicates that the α to β-PbO transformation between 489 °C and 620 °C was observed by several authors. The authors of this review note that this discrepancy is likely due to the slow kinetics of reaction [54]. Gavrichev *et al.* on the other hand pointed out later the influence of initial particles size on the reaction temperature, which they witnessed at 528 °C [53]. Real *et al.* note as well that lowering the initial particles size yields a higher β/α ratio [55]. It is yet unsure if an increase of the $O_2$ partial pressure favours the appearance of the α polymorph, as can be expected from its oxygen content [54], [55]. We will use 489 °C as the reaction temperature, as Risold *et al.* suggest. Lastly, the reverse (β to α) reaction was not often observed, probably due to the even slower kinetics of reaction [54]. The stability of PbO and its conversion in lead carbonates was studied in aqueous solution by Taylor and Lopata, who determined equilibrium conditions for each reaction in the PbO – plumbonacrite ($3PbCO_3 \cdot Pb(OH)_2 \cdot PbO$) – cerussite – hydrocerussite system [38]. Solid state carbonatation was less addressed. Clarke and Greene detected traces of hydrocerussite at the surface of PbO thin films heated at 176 °C for 96 h in air [39]. On the other hand, Mu *et al.* studied PbO nanoparticles dispersed in a silica gel and showed that they were transformed in $PbCO_3$ at ambient temperature and pressures between 1 and 15 bars. Moreover, they showed that water adsorbed at the nanoparticles surfaces played an important role in their carbonatation [40]. Lastly, note that PbO obtained from lead white irradiation by a pulsed Nd:YAG laser was shown to spontaneously reconvert to cerussite or hydrocerussite in air at ambient temperature in a few days [41]–[46].

---

[1] *Reconversion* is hereby used in the sense of a chemical reaction (conversion) taking place backward. Here the conversion is the oxidation to plattnerite; the reconversion is thus the formation of cerussite or hydrocerussite (white) or minium (red) from plattnerite (black) [5].

In this study, we investigated the conditions (temperature, atmosphere, time) in which white lead carbonates (cerussite, hydrocerussite and plumbonacrite) can be obtained from plattnerite. We first conducted an extensive preliminary study on powders, for analytical purposes. Then, guided by encouraging results, we demonstrated the possibility of obtaining lead carbonates by a thermal treatment of a plattnerite-based mural painting sample, which colour shifted from black to white.

## Materials and methods

This section outlines our experimental approach, as experimental details and procedures can be found in the SI section. For analytical purposes, we conducted most of the preliminary studies on raw plattnerite and massicot powders, which were heated in ovens and characterised by micro X-ray diffraction (µ-XRD) and scanning electron microscopy (SEM). We first investigated the conditions in which plattnerite is thermally decomposed in α-PbO and β-PbO, depending on the temperature and the heating time. To do this, a calibration curve linking µ-XRD α-PbO and β-PbO peaks areas to their mass fraction was set. Then we studied α-PbO and β-PbO behaviour in $CO_2$ and $CO_2/H_2O$ atmospheres at temperatures above 100 °C (by thermogravimetric analysis (TGA) and in an atmosphere-controlled oven) and below 100 °C (in sealed vials placed in an oven). Lead carbonates formation from massicot being observed below 100 °C, we made some full reconversion route trials on plattnerite based mural painting samples. The colour changes were monitored by spectrocolorimetry and photography, and XRD was used to assess the chemical transformation of plattnerite during the reconversion.

# Results and discussion

α-PbO synthesis from β-PbO$_2$. In order to produce α-PbO, we made a series of trials on β-PbO$_2$ powder using the Vulcan oven, between 480 °C and 590 °C by step of 10 °C, each lasting for several days (between 1 and 8). The resulting materials were characterised by µ-XRD (figure S-5 in the SI section) and the results are summarised in table 1. These data show that in order to obtain litharge, we must heat 5 g of plattnerite at 564 or 567 °C. In order to study the kinetics of litharge formation from plattnerite, we made a series of trials at 564 °C for less than a day. Results (figure S-6 in the SI section) show that plattnerite is converted in minium then litharge, and that minium disappear in *ca.* 5 to 7 hours. SEM observation showed that the facies does not vary much between 5 and 19 hours heating, but the mean grain size increases slightly (figure S-7 n the SI section).

Table 1: Oven litharge synthesis results from plattnerite

| Temperature | Results |
| --- | --- |
| 490 °C ⩽ T ⩽ 550 °C | Pure minium |
| 560 °C | Litharge with traces of minium |
| 564 °C, 567 °C | Litharge |
| 570 °C | Litharge with traces of massicot |
| 580, 590 °C | Litharge and massicot |

β-PbO synthesis from β-PbO$_2$. α to β-PbO transformation was observed at temperatures as high as 620 °C [54]. Thus, in order to produce β-PbO, we heated PbO$_2$ powder at 650 °C. µ-XRD of the resulting material showed that samples that spent little time (around and less than 1 h) at high temperatures exhibit traces of minium and plattnerite. We thus focused on samples that spent more time at high temperatures. We investigated the effect of the temperature rise and fall on the composition of the resulting powders by, first, producing two samples that spent 5 h at 650 °C before being quenched in air. One was heated up to 650 °C in 2 h and the other in 10 min. Their XRD diagrams were similar. Thus, we conclude that the heating rate had no effect on the resulting material in this time range. We then produced samples that spent 16 h at 650 °C: one was slowly taken down to ambient temperature in 4 h and the other was quenched in air and thus taken down to ambient temperature in *ca.* 10 s. Their XRD diagrams revealed a pure β-PbO composition for the latter, but the sample that had been slowly cooled exhibited traces of α-PbO, which is likely due to the reverse β to α reaction happening during the long cooling phase. The following samples were heated up in 10 min at 650 °C, spent various time at this temperature before being quenched in air down to ambient temperature in *ca.* 10 s. Their composition, assessed by μ-XRD, is pure PbO. Figure 1a shows their μ-XRD diagram in the highest intensity peaks of α and β-PbO range (respectively [1 0 1] α-PbO at 28.64° [56] and [1 1 1] β-PbO at 29.11° [57]). We can notice an increase of the [1 1 1] β-PbO to [1 0 1] α-PbO peak area ratio and that there is no trace of litharge for samples that were heated at 650 °C longer than 7 hours.

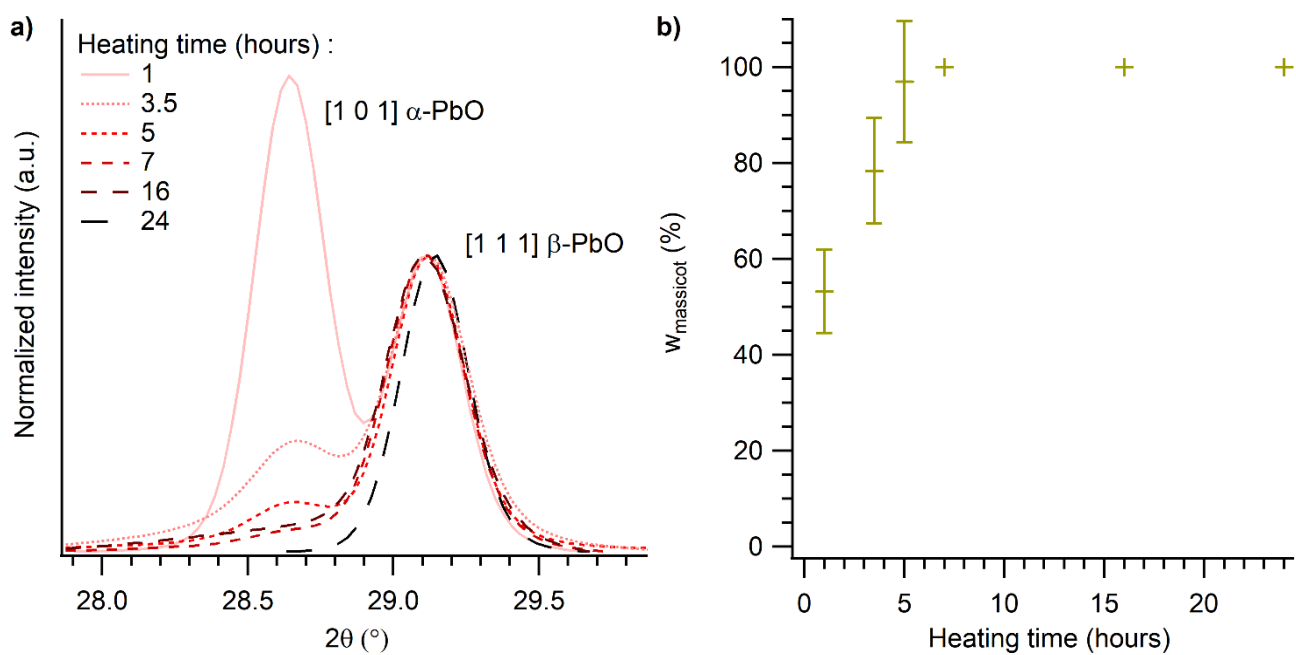

Figure 1: a) μ-XRD diagrams of β-PbO₂ powder heated up to 650 °C for increasing time and quenched in air down to ambient temperature in *ca.* 10 s. b) Pseudo β-PbO fraction, calculated using [1 1 1] β-PbO and [1 0 1] α-PbO fitting peaks areas.

Kinetics of the β-PbO synthesis from β-PbO₂. In order to link the relative α and β-PbO peak area to a mass ratio, we had to establish a calibration curve. We mixed pure α and β-PbO powders in various mass ratio and the resulting mixtures were characterised by μ-XRD. The main α and β-PbO peaks were fitted and a calibration curve was plotted (figure S-4 in the SI section), displaying the α to β-PbO mass ratio versus the [1 0 1] α-PbO to [1 1 1] β-PbO fitting peak area ratio. Likewise, the main α and β-PbO peaks displayed on figure 1a were fitted, and the β-PbO mass ratio was calculated using the calibration curve linear fit coefficients (figure 1b). It rises to 1 (pure massicot) linearly in 7 hours and remains constant thereafter. SEM images of the resulting powders (figures 2a and b) show rather plane crystals with 4 to 6 sides. Increasing the heating time seems to make most of the smallest crystals (around and less than a micron) disappear and to increase the average crystallite size. This was not seen on µ-XRD data, in which the fitting peaks FWHM did not vary with heating time (not displayed). Lastly, the crystals are also sharpened, with cleaner planes and edges when increasing the heating time.

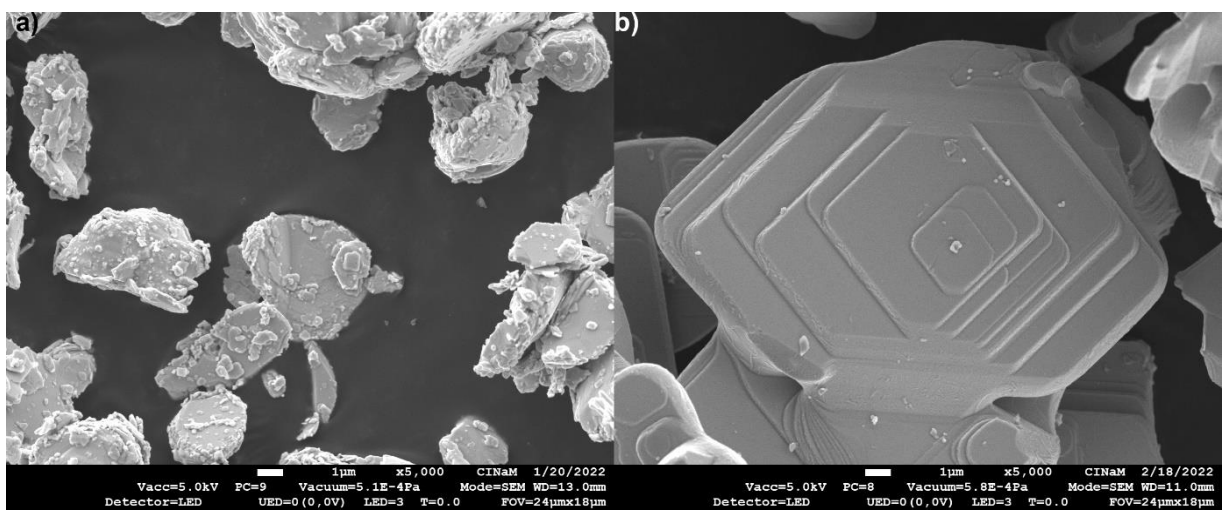

Figure 2: secondary electron SEM images of PbO powders having spent 5h (a, mixture of α and β-PbO) and 24h (b, pure β-PbO) at 650 °C.

β-PbO powder carbonatation at high temperature. The thermodynamic of PbO carbonatation was studied using the HSC chemistry software. No data were available for the hydrocerussite, so only the massicot to cerussite data were calculated. Figure 3 displays the calculated Gibbs free energy and enthalpy of reaction. The reactions are thermodynamically feasible when $\Delta G < 0$, thus under 300 °C. Moreover, the negative sign of $\Delta H$ in this temperature range indicates that this reaction is exothermic. At first, we tried to trigger the reaction at temperature as high as possible. We studied pure synthesised β-PbO powder, a synthesised PbO powder containing 54 % massicot (± 9 %), and a commercial PbO powder containing 83 % massicot (± 11 %), all of them with a *ca.* 10 µm granulometry. We heated the powders in a TGA device under $CO_2$ flux until 280 °C for 1h30 and in an oven under a bubbled $CO_2$ flux until 250 °C for 17h30. No reactions were detected on the TG signal during the analysis. No changes of colour were witnessed, and powder XRD diagrams were identical before and after the heating. We thus made the hypothesis that condensed water at the surface of PbO particles could play an important role in the carbonatation, as Mu *et al.* witnessed while studying PbO nanoparticles exposed to various pressures of dry and humid $CO_2$ at ambient temperature [40].

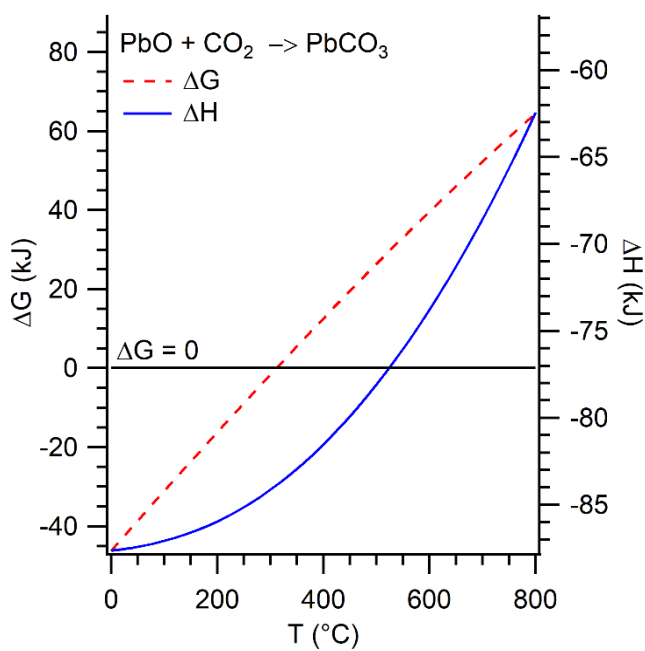

Figure 3: Gibbs free energy and enthalpy of the PbO to $PbCO_3$ reaction calculated with the HSC Chemistry software.

β-PbO powder carbonatation at low temperature. We thus studied the reaction for pure β-PbO powders in presence of various quantities of water, in $CO_2$ sealed vials, at temperatures lower than 100 °C. Note that the saturation vapor pressures, which we calculated using the Antisecos software [58], were always reached, even with the lowest quantity of water used. The effect of temperature and heating time were also studied and resulting powders were characterised by µ-DRX. In these conditions, a small fraction of the massicot was transformed in cerussite, hydrocerussite and plumbonacrite, another lead hydroxycarbonate, which is also white [59], [60]. It was already reported that plumbonacrite occurs as a degradation product in lead pigments-containing paintings [61], [62] or in the PbO-$CO_2$-$H_2O$ system in aqueous solution [38], [63]. It is reported to be metastable and to convert in hydrocerussite over time [63], [64]. The reactions progress between 1 and 7 days to a state that depends on the temperature and the quantity of water. Increasing the amount of water favours the appearance of plumbonacrite. Increasing the temperature favours the appearance of plumbonacrite too, and cerussite is not even present in powders that were heated up to 90 °C. We thus conclude that the carbonatation of massicot to cerussite, hydrocerussite and plumbonacrite is possible under 100 °C. The relatively low yield of the reaction could be explained by the relatively big size of the particles we studied, and the fact that it is raw powder, whereas Mu *et al.*, for example, focus their study on smaller nanoparticles dispersed in silica gel [40].

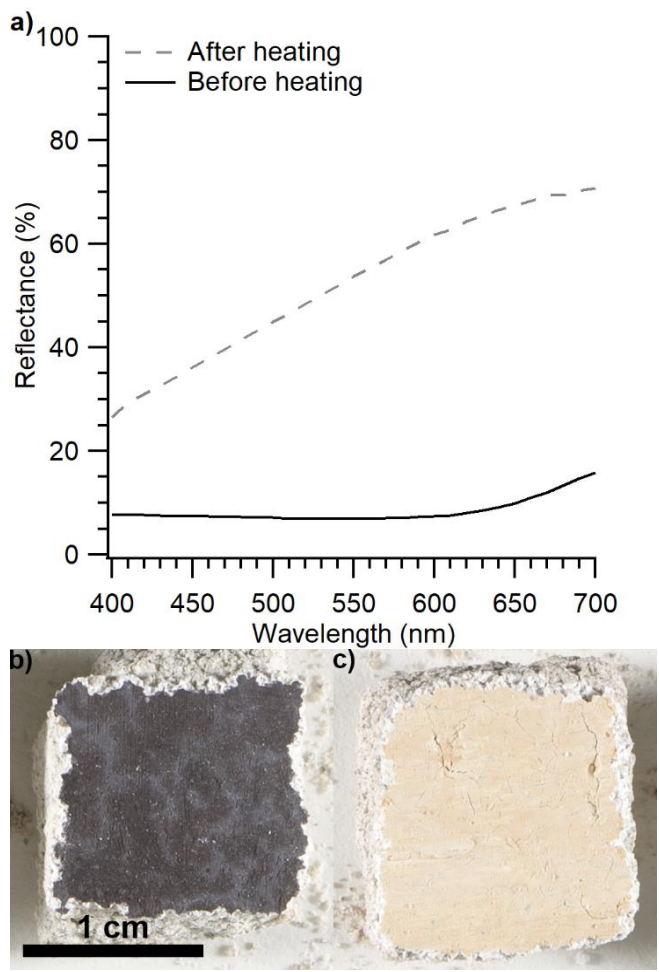

Figure 4: total reflectance spectra (a) and photos (b and c) of two plattnerite mural painting samples before (b) and after (c) 1 h heating at 650 °C.

Carbonatation tests on laboratory-made $PbO_2$ mural painting samples. Tests have thus been carried out on plattnerite-based lab-made mural painting samples. First, several fragments have been heated up to 650 °C in the Vulcan oven for various time and taken out. After 1 h heating, the resulting colour does not vary (figure S-8 in the SI section): however, their colour evolved from light yellow when just taken out of the oven to white/beige in an hour. Other fragments were cooled down with pressurised air for a few seconds. Their colours were then monitored by spectrocolorimetry and photography and their composition by XRD for 5 hours. No changes were observed, so the colour change that was observed is more likely to happen during the cooling phase, which is relatively long for a massive fragment that can store heat. Figure 4 presents photos of two samples: the initial unheated sample (figure 4b) and after 1 h heating at 650 °C (figure 4c), along with their total reflectance spectra (figure 4a). The colour shifted from black to white with strong red and yellow components, with a $\Delta E^* = 72.8$. The XRD analysis of the scrapped pictorial layers of these fragments showed that plattnerite was converted into massicot, plumbonacrite, but also lead silicate $Pb_3SiO_5$, and calcium lead oxide $Ca_2PbO_4$. These latter phases are probably the by-products of lead oxide phase's reaction with the calcite and quartz that were used for crafting the mural painting sample.

## Conclusion

Litharge and massicot were obtained by heating plattnerite powder for 7 hours respectively between 564 and 567 °C and at 650 °C. No changes were observed when heating massicot and litharge between 250 °C and 280 °C in dry or wet $CO_2$ up to a day. However, when heating massicot under 100 °C in a wet $CO_2$ atmosphere, part of it was converted in lead carbonates in a few hours: cerussite, hydrocerussite and plumbonacrite. Lastly, trials were made on plattnerite based mural painting samples at 650 °C for 1 hour, which yielded a mixture of massicot, plumbonacrite, but also lead silicate and calcium lead oxide, the latter probably being reaction products of lead species with calcite and quartz, which are present in the pictorial layer and mortar.

These results show that the thermal formation of white lead carbonates from plattnerite via a PbO intermediate is a viable reconversion route, although the formation of other lead species by reaction with mural painting constituents is to be considered. This makes the use of a continues wave laser for thermal reconversion of darkened lead white mural paintings a viable solution. At this point it would be interesting to do tests, to see if laser heating produces the same species as oven heating. Also, studying the reactions that happen at high temperature between plattnerite and some common pictorial layer and mortar constituent, such as lime, calcite and quartz would give insights on the phenomenon we observe. It would also help understanding how the heating and environmental parameters can be adjusted to influence the reaction towards the formation of cerussite and hydrocerussite rather than plumbonacrite and other lead species.

# Acknowledgement

The authors wish to acknowledge the French Ministry of Culture and Communication for financial support and the French Fondation des Sciences du Patrimoine for their financial support to the RECONVERT 2 project. The authors would also like to thank Kevin Ginestar from SCCME (CEA Saclay) for the help in the TGA and oven trials, Vasile Heresanu from CINaM (Aix-Marseille Université) for the µ-XRD experiments, Sébastien Aze (Sinopia) for helping in the preparation of mural painting samples, and Xueshi Bai for the discussion and advises throughout the RECONVERT project.

# Supplementary information

## Experimental section

Reference samples. Commercial plattnerite (97 % pure, Alfa Aesar A12742, lot number 10213539) and PbO (99.99 % pure, Alfa Aesar 14240, lot number X23F010) were used. SEM images of the plattnerite powder were taken, showing well defined submicrometric crystals (figure S-1)

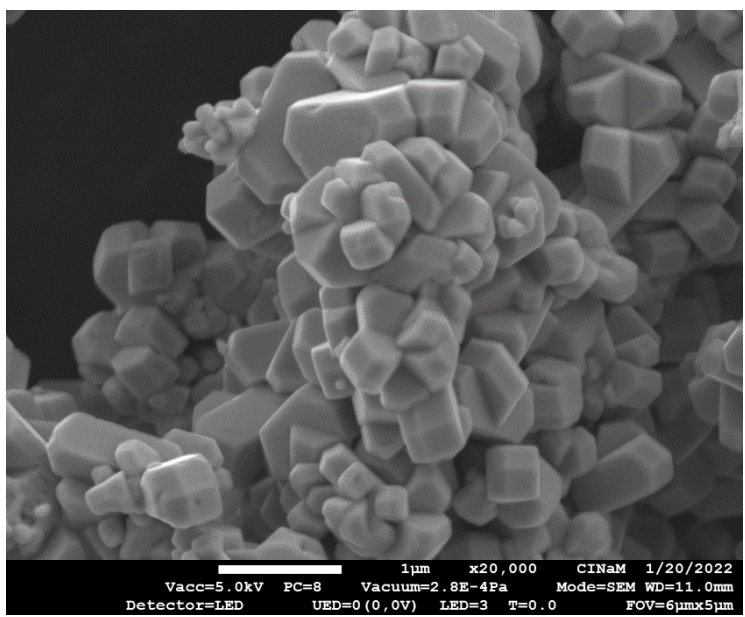

Figure S-1: secondary electron SEM images of the Alfa Aesar $PbO_2$ powder.

Mural painting samples. Mural painting samples have been made on Estaillades stone. Mortars were made by mixing coarse river sand 0/4 (Leroy Merlin), fine river sand 0/2 (Leroy Merlin), CL 90 air lime (Saint-Astier), NHL 3.5 hydraulic lime (Saint-Astier). Two plaster layers have been applied: a first 2 cm thick *arriccio* layer (3 volumes of coarse river sand, 1 volume of air lime), then a second 1 cm thick *intonaco* layer (2 volumes of fine river sand, 1 volume of hydraulic lime). Plattnerite was mixed with lime paste then applied on the plaster layers.

Vulcan oven trials. *Ca.* 5 g of β-$PbO_2$ were placed in an alumina crucible and heated in a Vulcan A-130 oven equipped with a temperature controller Eurotherm 3216. The temperature ramp rate cannot be controlled. The oven temperature rises to 650 °C in 10 minutes with an asymptotic profile (figure

S-2). When switched off, the temperature falls to ambient temperature in *ca.* 7 h (figure S-3). The powder is usually quenched in air and reach ambient temperature in *ca.* 10 s.

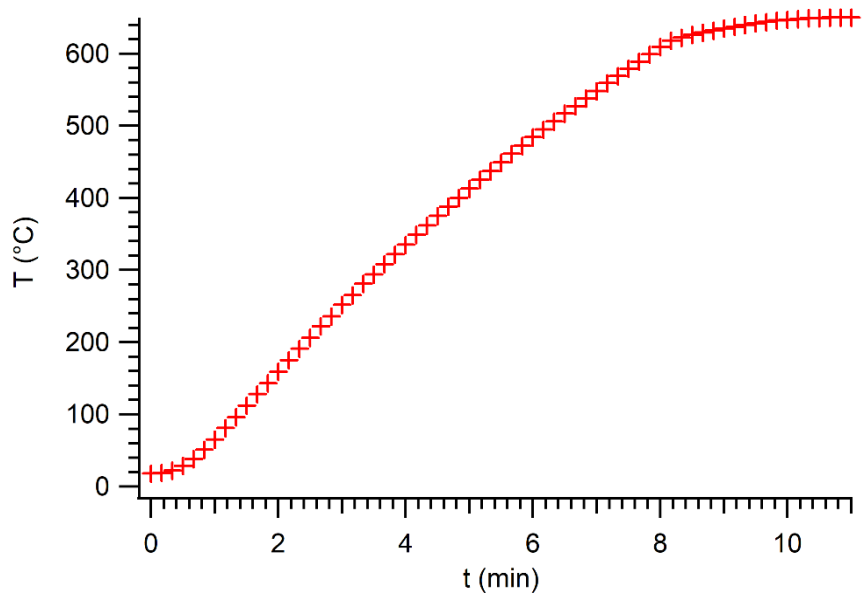

Figure S-2: temperature rise of the Vulcan A-130 oven when the temperature setpoint is set at 650 °C.

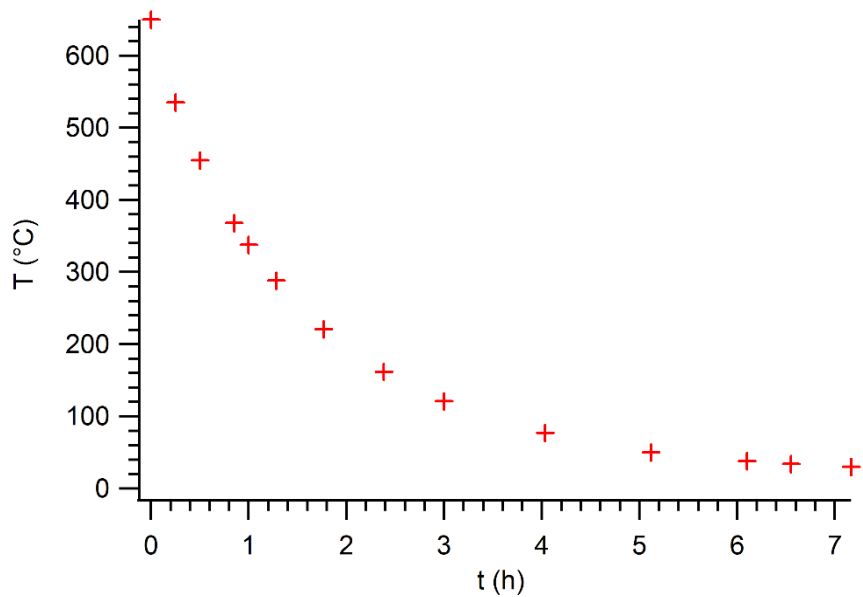

Figure S-3: temperature fall of the Vulcan A-130 from 650 °C, with door closed and no heating (outside temperature: 20 °C).

ΔG calculations. The Gibbs free energies of reaction were calculated with the HSC Chemistry software.

TGA. A Setaram B24 thermogravimetric analyser was used in the following conditions. We degassed the crucibles at 500 °C for a day before use. The sample crucible is usually loaded with *ca.* 140 mg PbO and the crucible usually weight *ca.* 325 mg. Before each trial stabilization of the system took place at 30 °C for 5 h. The crucibles are subjected to a 0.5 L min$^{-1}$ Ar flux and a 2.1 L min$^{-1}$ $CO_2$ flux. The mass change detection limit is equal to 10 µg in these conditions. A blank trial was first made with an empty crucible as a baseline. The temperature profile was as follow: heating up to 280 °C at 2 °C min$^{-1}$, followed by a 1h30 280 °C step, and sample cooling to ambient temperature at 20 °C min$^{-1}$. We studied pure synthesised β-PbO powder (from a 5 g batch of commercial β-$PbO_2$ that spent 24 h at 650 °C), a synthesised PbO powder mainly consisting in α-PbO (from a 5 g batch of commercial β-$PbO_2$ that spent 1 h at 650 °C then 24 h at 520 °C), and a commercial PbO powder consisting in a mixture of α and β-PbO. These powders where characterised by µ-XRD and SEM, which allowed to determine that their average granulometry is around 10 µm.

$CO_2$/$H_2O$ oven trials. We studied the same samples as in TGA in a tubular atmosphere-controlled oven. $CO_2$ was bubbled in water at ambient temperature (the mole fraction corresponding to saturation vapor pressures is 2.32 % at ambient temperature [58]) and then flowed at 15 L h$^{-1}$ in the quartz reactor where samples were placed in alumina vessels. The temperature profile was as follow: heating until 250 °C at 20 °C min$^{-1}$, followed by a 17h30 250 °C step, and sample cooling to 130 °C in 2 hours then the samples were taken out.

Sealed $CO_2$ vials experiment. In each 5 mL vial, 1 g of pure synthesised β-PbO powder was placed with various quantities of water (50 µL, 200µL, 3 mL), and $CO_2$ cartridges were opened and directed toward the vial before the latter was sealed. Vials were heated at 60 or 90 °C or left at ambient temperature for 5 h, 30 h or 8 days. After that, the powders were left to dry for 24 h at ambient temperature on watch glasses. Note that the water mole fractions corresponding to saturation vapour pressures are 2.32 % (ambient temperature), 19.8 % (60 °C) and 69.55 % (90 °C) [58].

XRD. A Bruker D8 Focus diffractometer has been used in the θ – 2θ mode, with a 0.0042864° step and 12 s/step. For XRD monitoring, we used a 0.0185744° step and 0.5 s/step. During the analysis, the sample spins at 15 rpm. The X-ray source is a Co anticathode (λ = 1.789 Å) operated in the following conditions: 40 keV / 35 mA.

µ-XRD. The XRD measurements were carried out using a device mounted on a high-gloss rotating Cu anode, Rigaku RU-200BH. A reflective focusing optic, OSMIC, mainly transmits the Cu Kα radiation (λ = 1.5418 Å) and a very small part of the Kβ, the latter being absorbed completely by a Ni filter. The detector used is 2D, flat image type, model Mar345. For these measurements, the working power was 50 kV and 50 mA and the beam size was $0.5 \times 0.5$ mm$^2$. XRD patterns were collected from quartz capillary containing crushed samples.

µ-XRD data treatment. Background was linearly fitted between 22.99° and 25.986° (in a region where no α-PbO or β-PbO or $Pb_3O_4$ peak are present) then subtracted to the data. Normalization was made on the highest intensity point of [1 1 1] β-PbO peak, around 29.11° [57]. Data were fitted by a linear baseline and two Voigt function between 27.879° and 29.886°, covering the [1 0 1] α-PbO and [1 1 1] β-PbO range (respectively at 28.64° [56] and at 29.11° [57]). The errors on the fitting peaks areas being usually lower than 0.002, thus yielding a pseudo massicot fraction error lower than 0.01.

Calibration curve establishment and use. 11 mixtures of massicot and litharge were prepared, each representing around 100 mg massicot. An average 0.56 % error was estimated when emptying the dish after weighting, which thus represent the massicot fraction error. Mixtures were characterised by µ-XRD and the fitting peak ratio were calculated and plotted (figure S-4). The curve was linearly fitted (y = a + b*x), with resulting coefficient being a = 11.281 ± 4.85 and b = 103.26 ± 9.45. The fitting $R^2$ was 0.922667. We then used this fitting coefficients in order to calculate the massicot fraction from the fitting peak area ratios of mixtures of unknown massicot fractions. The extrapolated massicot fraction error was calculated from the relative error on a and b coefficients.

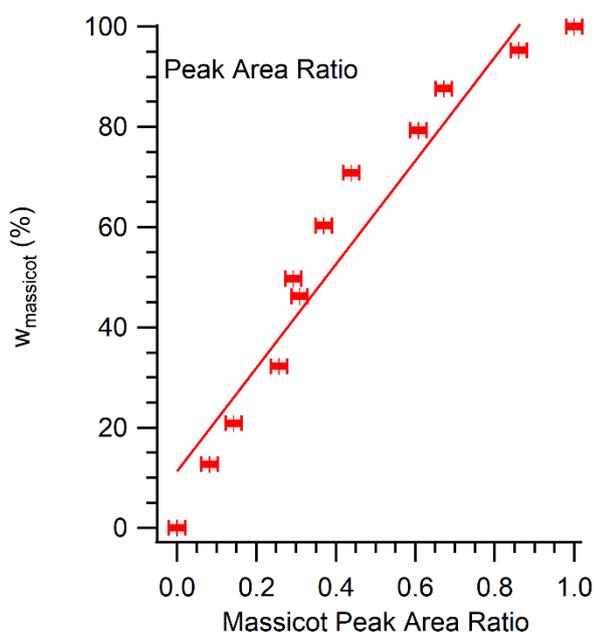

Figure S-4: massicot/litharge μ-XRD calibration curve.

SEM-EDS. The microscope is a JEOL JSM-7900F usually operating at 5 kV for imaging purposes at a working distance of 11 to 13 mm, equipped with an in-chamber LED secondary electron detector. For EDS analysis with the Bruker annular four channel SDD Flat-Quad detector, the microscope was operated at 15 kV at a 12.5 mm working distance. Spectra were acquired until reaching one million counts, which takes around one minute.

Spectrocolorimetry. We used a Hunter Lab MiniScan XE Plus spectrocolorimeter operated by the Universal Software. We used a D65 illuminant and a 10° observer. Three visible spectra are acquired in a few seconds and averaged; the resulting L*a*b* values are calculated by the software.

Photography. A Nikon D80 with a Nikon AF-S Micro NIKKOR 60mm f/2.8G ED objective equipped with two Elinchrom D-Lite RX2 flashlights was used at *ca.* 30 cm from the sample in the following conditions: ISO 100, aperture f/11, 1/125 s shutter speed, 2.9 flash power. A Calibrite colorchecker passeport photo 2 colour chart was used for the white balance on the black, third gray and white squares with the following RGB values : (53; 53; 53) (125; 125; 125) (243; 243; 243).

# Supplementary material

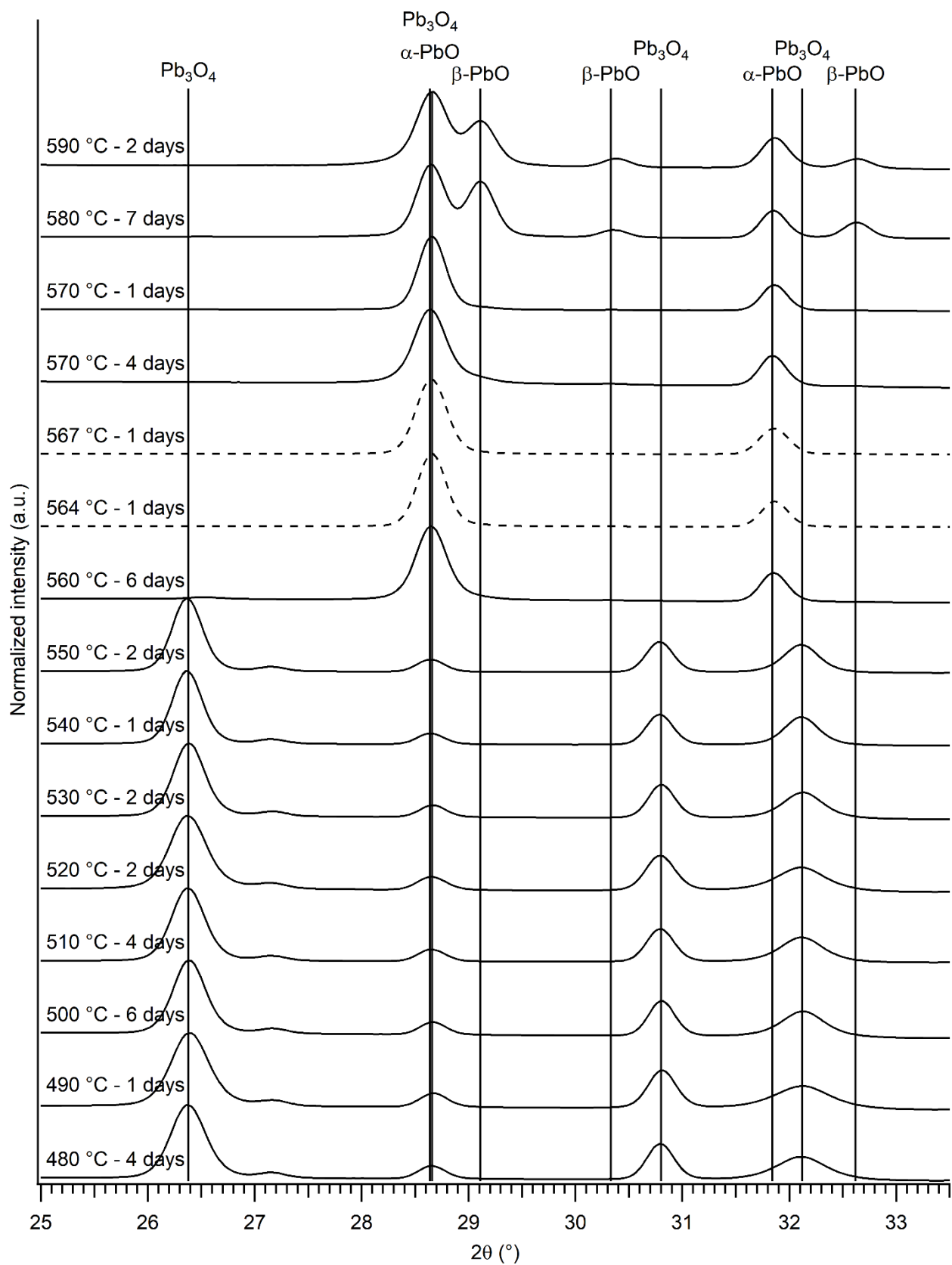

Figure S-5: μ-XRD diagrams of the resulting materials from the series of trials on β-PbO₂ powder using the Vulcan oven. Linear backgrounds were subtracted between 22° and 24° and the diagrams

were normalised on the highest intensity peak in the [25°; 34°] region and shifted for visibility. In dashed lines are the two trials yielding pure litharge.

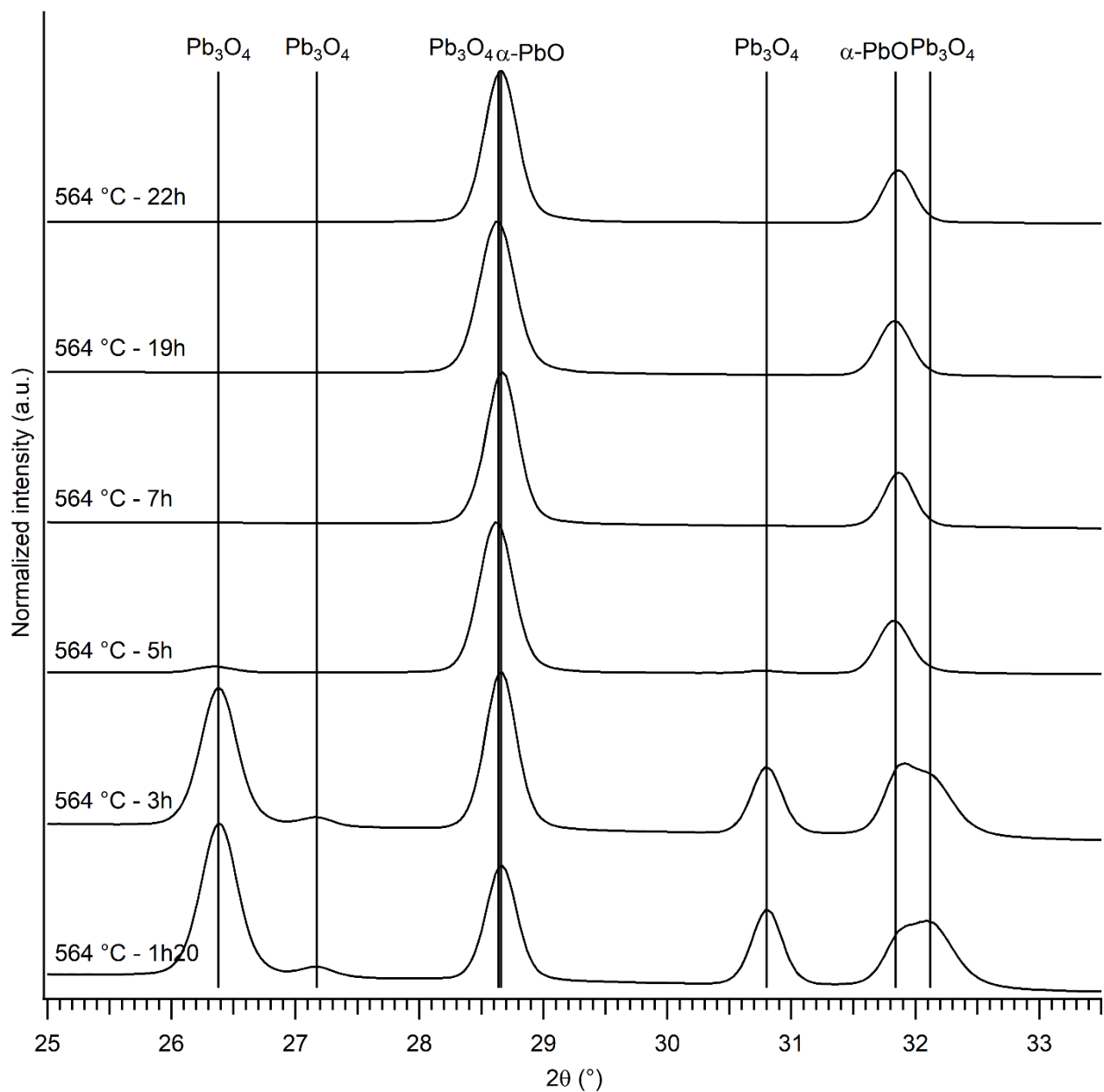

Figure S-6: µ-XRD diagrams of the resulting materials from the series of trials on β-$PbO_2$ powder using the Vulcan oven. Linear backgrounds were subtracted between 22° and 24° and the diagrams were normalised on the highest intensity peak in the [25°; 34°] region and shifted for visibility.

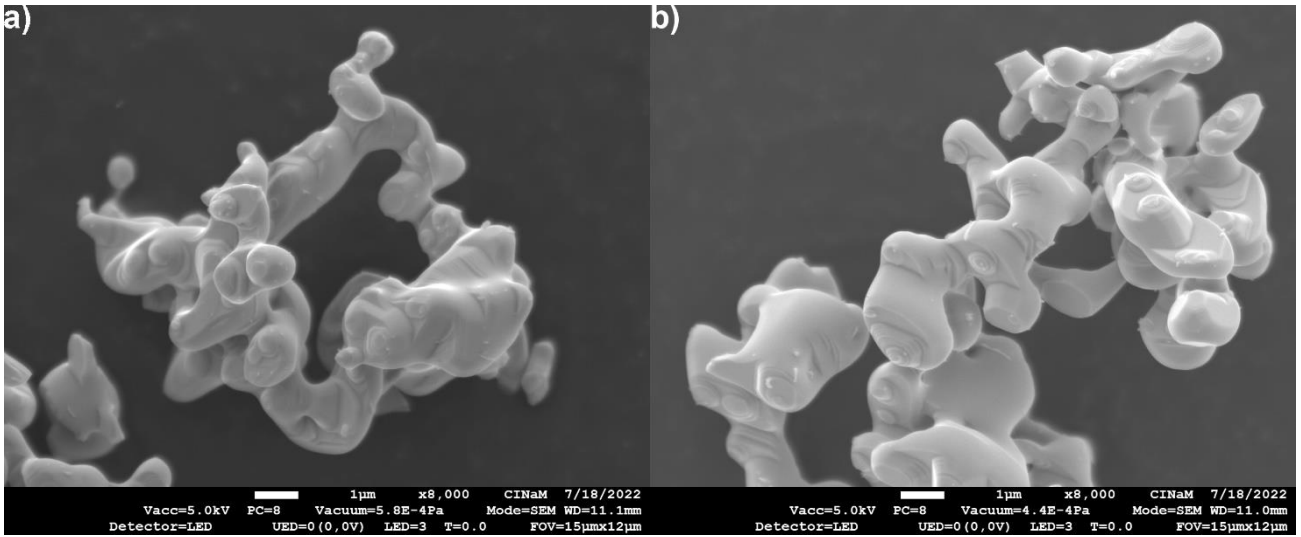

Figure S-7: secondary electron SEM images of α-PbO powders having spent 5h (a) and 19h (b) at 564 °C.

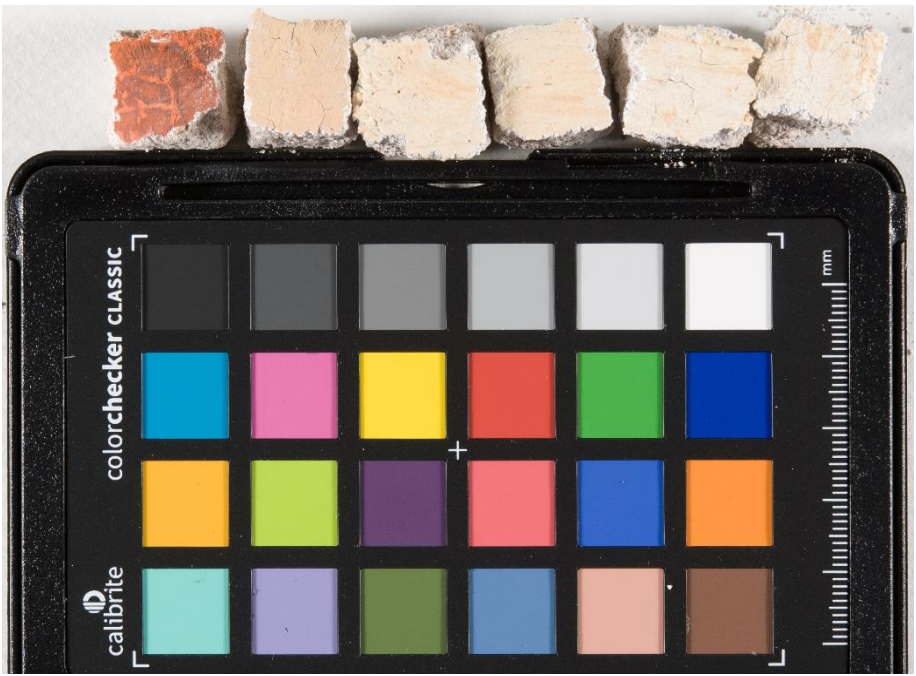

Figure S-8: plattnerite-based laboratory-made mural painting samples that were heated up to 650°C in 10 min and taken out of the oven after: 0 min, 30 min, 60 min, 90 min, 3 h and 5 h 30 min (from left to right) and left 1 day at ambient temperature and atmosphere.